\documentclass[10pt]{article}
\usepackage{amsmath,amssymb,amsthm,epsfig,graphicx}
\newcommand{\e}{{\bf e}}

\newcommand{\beq}{\begin{equation}}
\newcommand{\Tfour}[5]{
  \begin{picture}(40,35)(0,0)
  \put(-2,20){\psfig{figure=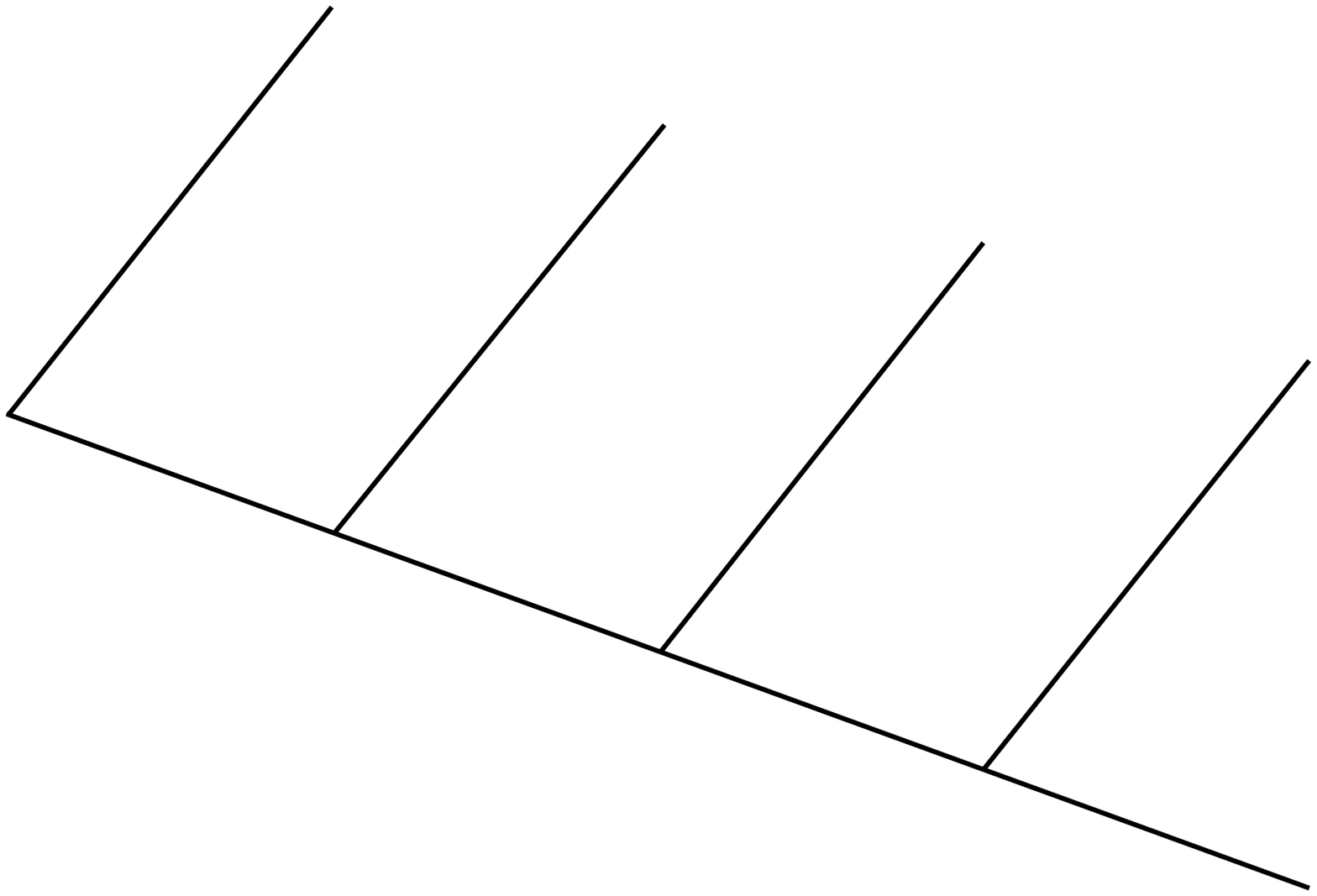,width=30pt,angle=-90}}
  \put(27,10){$\scriptstyle  #1$}
  \put(24,2){$\scriptstyle   #2$}
  \put(21,-6){$\scriptstyle  #3$}
  \put(18,-14){$\scriptstyle #4$}
  \put(0,-14){$\scriptstyle  #5$}
   \end{picture}}
\newcommand{\Tthree}[4]{
  \begin{picture}(40,30)(0,0)
  \put(-2,20){\psfig{figure=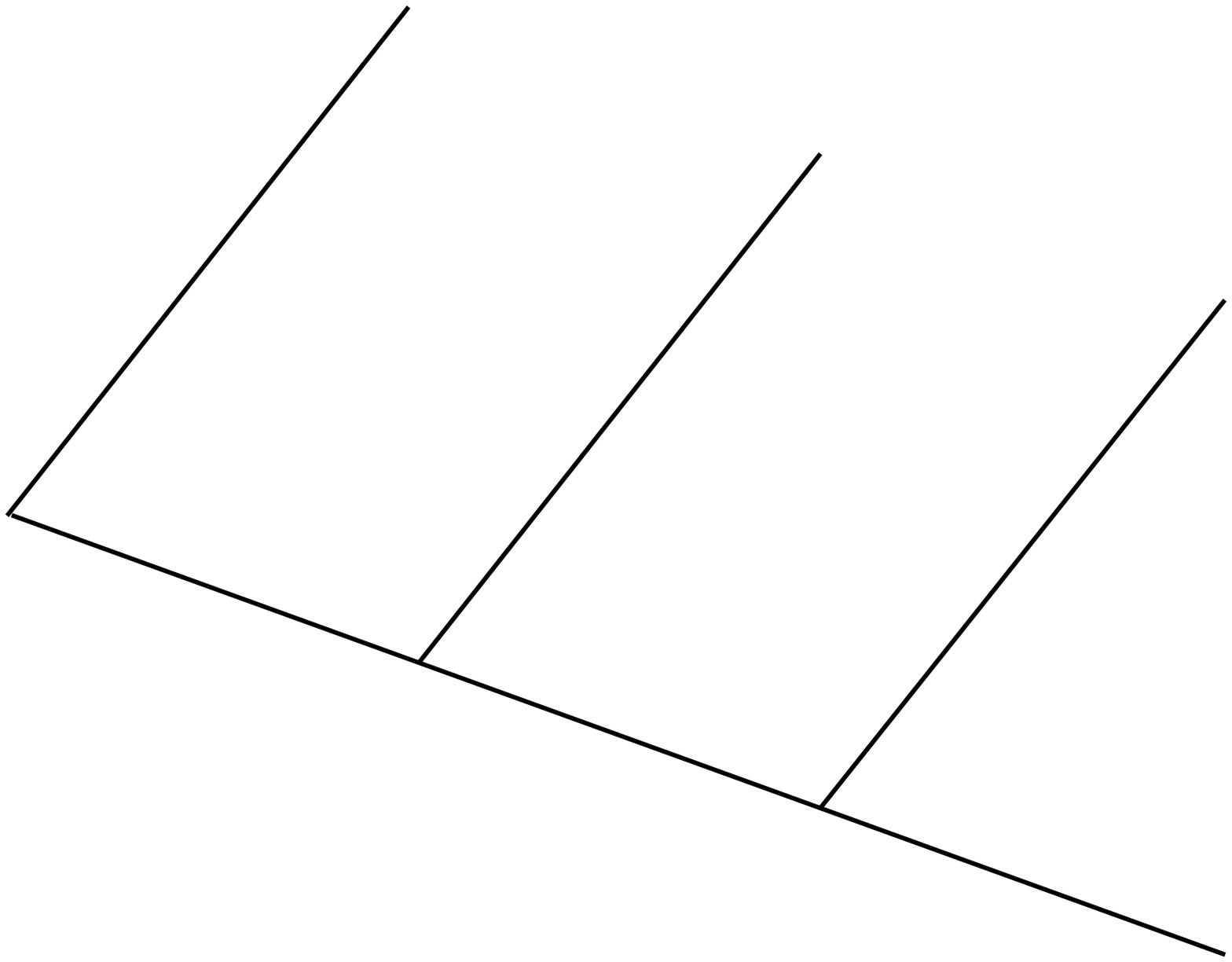,width=30pt,angle=-90}}
  \put(28,8){$\scriptstyle   #1$}
  \put(25,-3){$\scriptstyle  #2$}
  \put(22,-14){$\scriptstyle #3$}
  \put(0,-14){$\scriptstyle  #4$}
   \end{picture}}
\newcommand{\Ttwo}[3]{
  \begin{picture}(40,35)(0,0)
  \put(-2,17){\psfig{figure=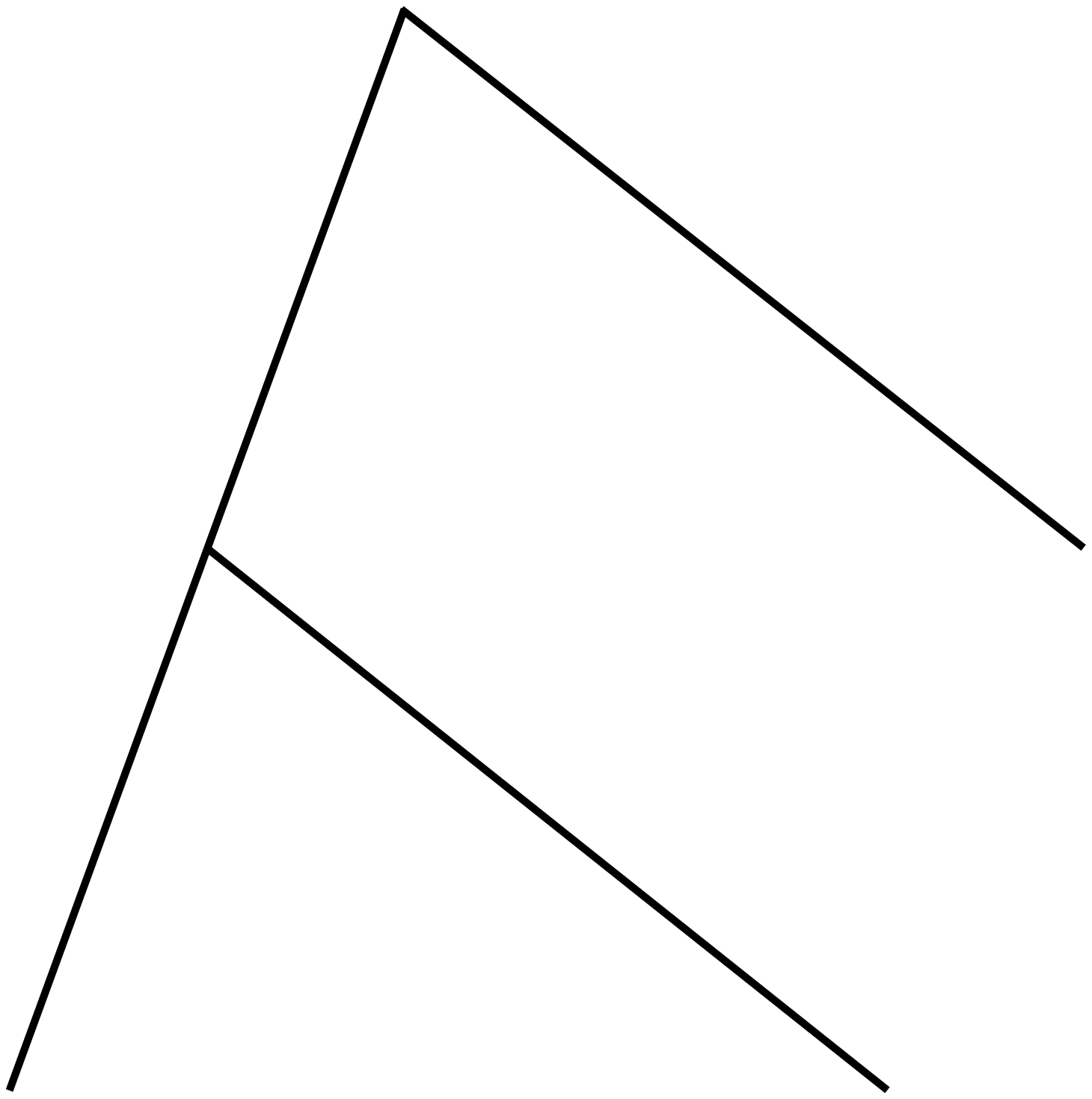,width=30pt,angle=-90}}
  \put(28,0){$\scriptstyle   #1$}
  \put(22,-14){$\scriptstyle  #2$}
  \put(0,-14){$\scriptstyle  #3$}
   \end{picture}}
\newcommand{\Tone}[2]{
  \begin{picture}(25,35)(0,0)
  \put(-5,17){\psfig{figure=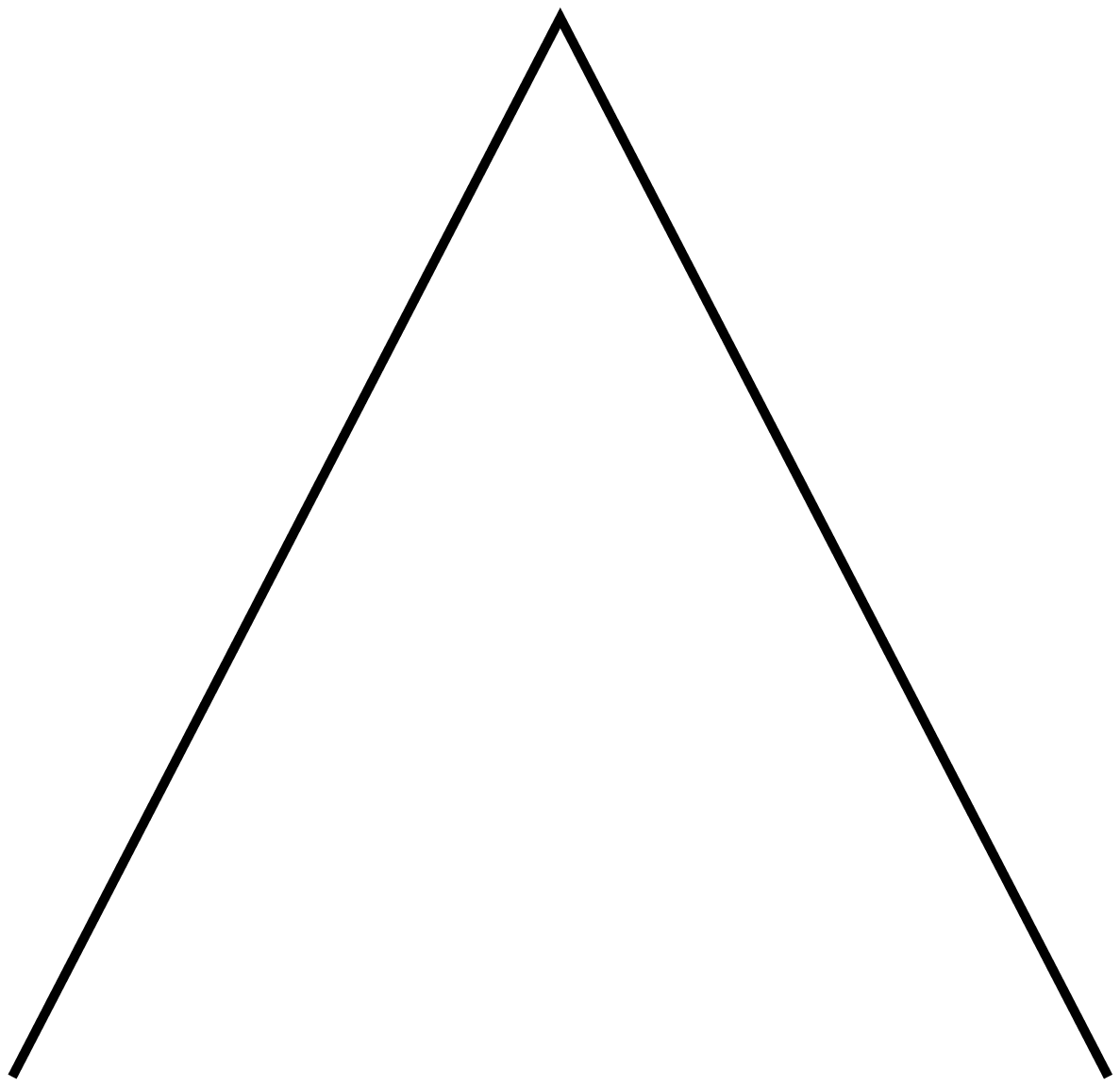,width=30pt,angle=-90}}
  \put(21,-14){$\scriptstyle   #1$}
  \put(-1,-14){$\scriptstyle   #2$}
   \end{picture}}
\theoremstyle{plain}

\theoremstyle{definition}

\theoremstyle{remark}

\begin{document}
\begin{titlepage}
\renewcommand{\thefootnote}{\fnsymbol{footnote}}
\begin{flushright}
ICN--2002--100, \today
\end{flushright}
\vspace{0.1in}
\LARGE
\center{HOPF ALGEBRA PRIMITIVES IN PERTURBATION QUANTUM FIELD THEORY}
\vspace{0.2in}
\center{M. Rosenbaum\footnote{E-mail: {\tt mrosen@nuclecu.unam.mx}}, J. David Vergara\footnote{E-mail: {\tt vergara@nuclecu.unam.mx}}
and H. Quevedo\footnote{E-mail: {\tt quevedo@nuclecu.unam.mx}}}
\center{Instituto de Ciencias Nucleares, \\
Universidad Nacional Aut\'onoma de M\'exico, \\A. Postal 70-543, \\ M\'exico D.F., M\'exico.}
\normalsize
\begin{abstract}
 The analysis of the combinatorics resulting from the perturbative expansion of the
 transition amplitude in quantum field theories, and the relation of this expansion
 to the Hausdorff series leads naturally to consider an infinite dimensional Lie
 subalgebra and the corresponding enveloping Hopf algebra, to which the elements
 of this series are associated. We show that in the context of these structures the
 power sum symmetric functionals of the perturbative expansion are Hopf primitives
 and that they are given by linear combinations of Hall polynomials, or
 diagrammatically by Hall trees. We show that each Hall tree corresponds to sums
 of Feynman diagrams each with the same number of vertices, external legs and loops.
 In addition, since the Lie subalgebra admits a derivation endomorphism, we also show
 that with respect to it these primitives
are cyclic vectors generated by the free propagator, and thus
provide a recursion relation by means of which the (n+1)-vertex
connected Green functions can be derived systematically from the
n-vertex ones. 


\end{abstract}
\end{titlepage}

\setcounter{footnote}{0}

\section{Introduction}

It has been conjectured \cite{conkre3} that
because of the interplay of Quantum Mechanics and General 
Relativity at the Planck scale, space-time ought to be regarded 
as a derived concept whose
structure should follow from the properties of Quantum Field Theory.
Perturbative
quantum field theory (PQFT), albeit its conceivable limitations
at distances of the order of the Planck length, is the only computational
tool available at present, and so the elucidation of the
mathematical structures behind that theory is a suggestive first step in 
pursuing this line of thought. One of these structures, which is based on the original work of Kreimer \cite{kreimer2} and further developed in \cite{kreimer1, kreimer2001, conkre1, conkre2}, is now generally known as the Hopf Algebra of Renormalization and it provides the underlying mathematics behind the
Forest Formula in the process of renormalization. Basically this Hopf algebra can be
represented by Feynman diagrams or decorated rooted
trees, where decorations are one-particle irreducible (1PI)
divergent diagrams without subdivergences. Other Hopf algebras
related to rooted trees and
to the Hopf Algebra of Renormalization have been discussed in the
literature, such as the vector space Hopf algebra of rooted trees
of Larson and Grossman \cite{gross}. The connection of this
algebra with the algebra of Kreimer and Connes was analyzed in
\cite{pana} and more recently revised in \cite{hoff}. For a
formulation in terms of a single mathematical construction of 
the several Hopf algebras described by rooted
trees see van
der Laan \cite{laan}.\\

The essential point of the Kreimer-Connes formalism is that {\bf
given that a theory is renormalizable} an appropriately defined
twisted antipode, based on the minimal subtraction scheme of
renormalization, generates the counterterms corresponding to the
BPHZ Forest Formula and the antipode axiom provides a systematic
procedure for deriving the physically correct and finite
expression for a given diagram. However, the mere fact that the
twisted antipode axiom, or for that matter the Forest Formula,
provide a finite answer does not suffice to make the theory
physical. It is crucial, in order that the theory be
renormalizable, that the resulting counterterms are of the same
form as those in the original Lagrangian and that they can be
absorbed into the bare parameters of the renormalized Lagrangian
in a consistent manner. Generally this is not possible, and in such a
case the theory is described as non-renormalizable. Of course if
we know {\it a priori} that the theory is renormalizable then the
Hopf algebra of decorated rooted trees or of Feynman diagrams
remains most valuable both as the mathematical structure behind
the Forest Formula as well as for a systematic construction
of renormalized Green functions.\\

Another Hopf algebra related to rooted trees by a canonical mapping
is the algebra of normal coordinates that was recently
discussed in \cite{chryss}. In that work undecorated rooted trees
were considered, and the relevance of normal coordinates to the concept of
k-primitiveness as well as their role in the process of renormalization was analyzed. 
It was also shown there that for undecorated ladder trees, or to that
effect for non-branched trees with only one decoration (such as in rainbow
diagrams), the renormalization of the associated normal
coordinates is a one step procedure. However, when the diagrams
for a theory involve more than one decoration (as is usually the
case) ladder normal coordinates are in general no longer primitive,
even though one can still expect them to posses a milder pole
structure than that present in the rooted tree coordinates. 
Some of the pertinent questions left open in the above cited paper
concerned the physical interpretation of the normal coordinates
and whether perturbation theory could be formulated directly in terms
of them without having to go first through the algebra of rooted trees
or Feynman diagrams.
These questions provided some partial motivation for the present work
where we investigate some additional Hopf algebra structures
that can be associated naturally to the Hausdorff series expansion of PQFT.
Specifically, we show here that the power 
sum symmetric functionals of these perturbative expansions are elements of 
a free Lie subalgebra as well as Hopf primitives of its enveloping free algebra. Moreover
these primitives can be expressed as linear combinations of Hall polynomials
and, diagrammatically, as Hall trees. We show also that a Hall
tree corresponds to sums of Feynman diagrams, each with the same number of vertices,
external legs and loops. \\

In addition, since the Lie subalgebra admits a derivation endomorphism, we further show
that - with respect to it - these primitives
are cyclic vectors generated by the free propagator, and thus
provide a recursion relation by means of which the (n+1)-vertex
connected Green functions can be derived systematically from the
n-vertex ones. Lastly we show that the Hopf primitives considered here are normal coordinates resulting from a canonical mapping applied to the Poincar{\'e}-Birkhoff-Witt basis constructed from the complete homogeneous symmetric functionals, which appear also naturally in the Hausdorff series expansion of PQFT. This combinatorics is reminiscent of the one appearing in \cite{chryss}, but the possibility of a deeper relationship between the two Hopf algebras, and therefore a possible physical interpretation for the normal coordinates in \cite{chryss}, requires further investigation.\\

The paper is structured as follows: In section 2 we begin with a brief review of
the main steps that lead to the
perturbative expansion of the transition amplitude, and describe its relation to
the Hausdorff series and to the associated Hopf algebras for which the power sum functionals
are primitives. In subsections 2.1.1-2.1.3 the free algebra related to these
structures is discussed and it is shown that the functional primitives are linear combinations of the Hall polynomials that generate the Lie subalgebra of our free algebra. We also show that this Lie subalgebra admits a derivation endomorphism with respect to which all the primitives are cyclic vectors generated by the free propagator. The diagrammatic representation of the primitives in terms of Hall trees is also discussed in this section and shown to give a clear image of this cyclicity and of the iteration process by means of which the $(n+1)$-vertex connected Green functions can be constructed from
the $n$-vertex ones. Section 3 is devoted to a discussion of our main results and possible lines of future related research. We also give there an explicit and heuristic argument, based on a Birkhoff decomposition of the Hopf algebra considered here, which we believe helps to stress some of the points made in this Introduction.

\section{Algebraic Structures in Perturbation Quantum Field Theory}

Let us begin with a brief summary of the essential steps in PQFT
leading to the Green functions, with the dual purpose of making our presentation
self contained as well as for identifying the basic mathematical and physical
entities to which the Hopf algebras considered here are related.\\

For an arbitrary field theory the Euclidean
transition amplitude (the formulation in Minkowski space is achieved by analytic continuation) is given by
\begin{equation}
W_{E}[{\bf J}]=N\int {\mathcal D}{\bf \Phi}\;  \e^{-\int d^{d} x
[{\cal L}_0+ {\cal L}_{int} -{\bf J\cdot \Phi}]}. \label{01}
\end{equation}
Here $\Phi$ denotes the set of fields appearing in the theory, and
${\bf J}$ denotes the set of arbitrary currents introduced to drive each
field\footnote{Following Ref.\cite{ramond} we use the notation $W_E[J]$
for the generating functional of connected and disconnected graphs, and  
$Z_E[J]$ for the generating functional of connected graphs, sometimes the opposite 
notation is used in other works.}.
Using functional derivatives the amplitude (\ref{01}) can be rewritten as
\begin{equation}
W_{E}[{\bf J}]=\e^{-\langle{\cal
L}_{int}\left(\frac{\delta}{\delta {{\bf J}_{x}}}\right)\rangle
_{x}} \e^{-Z^{0}[{\bf J}]} W_{0}[0] \label{02}
\end{equation}
where ${\cal L}_{int}\left(\frac{\delta}{\delta {{\bf
J}_{x}}}\right)$ is the Lagrangian of interaction written in terms
of functional derivatives of the field currents ${\bf J}(x)$, and
$W_0[{\bf J}]=  \e^{-Z^{0}[{\bf J}]}W_0[0]$ is the free generating
functional. The symbol $\langle\;\rangle_{x}$ stands for
integration over the variable $x$ (after acting to the right with
the functional derivative). Note that the functional derivatives
with respect to the currents act according to the Leibnitz rule on
the term $\e^{-Z^{0}[{\bf J}]}$, and functional derivatives that
go through to the right of that term cancel
when acting on $W_{0}[0]$. Thus here $W_{E}[{\bf J}]$ is a functional and not an operator.\\

To simplify our exposition we shall consider the neutral
scalar $\varphi^{4}$ theory in Euclidean 4-dimensions when doing
explicit calculations. It is well known that this theory is
renormalizable and a clear discussion of the steps leading to its
renormalization may be found in \cite{ramond}. For this case  the transition
amplitude in (\ref{01}) reduces to
\begin{equation}
W_{E}[J]=N\int {\mathcal D}\varphi\; \e^{-\int d^{4} x [\frac{1}{2}\partial_{\mu} \varphi
\partial_{\mu} \varphi +
\frac{1}{2} m^{2}
\varphi^{2}+V(\varphi)-J\varphi]}, \label{1b}
\end{equation}
and (\ref{02}) becomes
\begin{equation}
W_{E}[J]= \e^{-\langle V(\frac{\delta}{\delta J_{x}})\rangle
_{x}} e^{-Z^{0}[J]} W_{0}[0], \label{1}
\end{equation}
where
\begin{equation}
\langle V(\frac{\delta}{\delta J_{x}})\rangle  _{x}=\int d^{4}\;x
\; \frac{\lambda}{4!} \frac{\delta^{4}}{\delta J_{x}^{4}},
\label{1a}
\end{equation}
\begin{equation}
W_{0}[0]= N\int {\mathcal D}\varphi\; \e^{-\int d^{4} x [\frac{1}{2}\partial_{\mu} \varphi
\partial_{\mu} \varphi + \frac{1}{2} m^{2}
\varphi^{2}]}, \label{2}
\end{equation}
\begin{equation}
 Z^{0}[J]=\frac{1}{2}\langle J(x)\Delta_{xy} J(y)\rangle  _{xy},\label{3}
\end{equation}
and
\begin{equation}
\Delta_{xy}=\frac{1}{(2\pi)^{4}}\int d^{4}p \frac{\e^{ip\cdot(x-y)}}{p^{2} + m^{2}}\label{4}
\end{equation}
is the Feynman propagator in 4-dimensional Euclidean space.\\

Writing $W_{E}[J]=\e^{-Z_{E}[J]}$ and rearranging (\ref{1})
results in
\begin{equation}
Z_{E}[J]= - \ln W_{0}[0] + Z^{0}[J] - \ln \left(1 + \e^{Z^{0}[J]}
\left(e^{-\langle V(\frac{\delta}{\delta J})\rangle
}-1\right)\;\e^{-Z^{0}[J]}(1)\right),\label{5}
\end{equation}
where, in order to make both sides of the above equation
consistent, we have explicitly included the action on the identity
function of the operator inside the logarithm, so as to cancel
derivations to the right of $\e^{-Z^{0}[J]}$. Now let
\begin{equation}
\sigma(\lambda):=\e^{Z^{0}[J]}\left(\e^{-\langle
V(\frac{\delta}{\delta J})\rangle
}-1\right)\;\e^{-Z^{0}[J]}(1)=\sum_{k\geq 1} \lambda^{k}S_{k}[J],
\label{6}
\end{equation}
where the second equality is the formal power series expansion of the first one in terms of
the coupling constant $\lambda$ in the potential.
If we further define
\begin{equation}
\psi(\lambda):=\sum_{k\geq 1} \lambda^{k}\psi_{k}[J]=\ln (1 + \sum_{k\geq 1} \lambda^{k}S_{k}[J]),\label{7}
\end{equation}
it then readily follows from (\ref{5}) that
\begin{equation}
Z_{E}[J]= -\ln W_{0}[0]+ Z^{0}[J] - \sum_{k\geq 1} \lambda^{k} \psi_{k}[J].\label{8}
\end{equation}

Note that in the theory of symmetric functions \cite{mac,gelfand} an
expression like (\ref{7}) relates the complete homogeneous symmetric functions to
the so called power sum symmetric functions or Schur polynomials, so we can use the
combinatorics of that theory
to write down the explicit (invertible) relation between the functionals $\psi_{k}[J]$ and the $S_{i}[J]$,
as homogeneous polynomials of order $k$ in the latter. This is given by
\begin{equation}
\psi_{k}[J]=\sum_{|I|=k} (-1)^{l(I)-1}\;\frac{S^I}{l(I)}, \label{11}
\end{equation}
where $I$ is a composition $I = (i_{1},\dots, i_{r})$ of
nonnegative integers, $|I|= \sum_{k} i_{k}$ is its weight,
$l(I)=r$ is its length and $S^I= S_{i_{1}}S_{i_{2}}\dots
S_{i_{r}}$.
In terms of the Feynman diagrammatic representation, the $\psi_{k}[J]$ correspond to linear
combinations of connected graphs each with $k$ vertices. \\

Defining the operator
\begin{eqnarray}
X&=&-\frac{1}{\lambda} \e^{Z^{0}[J]}\left(\langle
V(\frac{\delta}{\delta J_{x_1}})\rangle  _{x_1}\right)
\;\e^{-Z^{0}[J]}\nonumber\\
 &=& -\frac{1}{\lambda}\sum_{n\geq 0}\frac{1}{n!} \text {ad}(Z^{0}[J])^{n}
 \left(\langle V(\frac{\delta}{\delta J_{x_1}})\rangle  _{x_1}\right),\label{9}
\end{eqnarray}
where the adjoint operator in the second equality is defined as
the right normed bracketing: $\text {ad}(b)^{n}(a)\equiv
[b,[b,\dots, [b,a]]\dots ]$, one can verify that the functionals
$S_{k}$ introduced above are given by the recursion relations
\begin{eqnarray}
S_{1}[J]&=& X(1)\nonumber \\
S_{n}[J]&=&\frac{1}{n}X(S_{n-1}[J])(1),\label{10}
\end{eqnarray}
where $X(1)$ denotes the action of the operator (\ref{9}) on the
identity function. \\

In the same way that $X$ generates a recursion relation for the functionals $S_{k}$,
we will show bellow that a derivation operator can be defined which when acting on the
$\psi_{k}[J]$ leads to expressions analogous to (\ref{10}). We first analyze some of the
algebraic structures behind the operators occurring in (\ref{1}) and (\ref{5}).
\\

\subsection{Algebraic Structures}

As mentioned in the Introduction, one of our goals in this paper is to
investigate the relevance to the process of renormalization of
Hopf algebra primitives that occur naturally at a prediagram stage
in PQFT. Therefore these primitives must be related to the two entities appearing
in the perturbation expansion of the transition amplitude, {\it i.e.} the complete
homogeneous symmetric functionals and the power sum symmetric functionals.

From the theory of symmetric functions we know that one can
associate respective Hopf algebra structures to these two sets of
functionals for which the $\psi_{k}[J]$ are primitives. In
addition to the above, there is a free algebra generated
by the operator $\langle V \rangle $ and the functional $Z^{0}[J]$
for which the $\psi_{k}[J]$ are also primitives. We describe these
three Hopf algebras which are related by the following diagram

\[
\begin{array}{cccc}
K< \left\lbrace S_{k}\right\rbrace> \simeq K< \left\lbrace \psi_{k}\right\rbrace > & \subset K< \left\lbrace
\Psi_{k}\right\rbrace > & \hookrightarrow & K < Y >
\\
&\bigcup & &    \bigcup \\
& {\frak L}_{K}(\{\Psi_{k}\}) & \hookrightarrow & {\frak L}_{K}(Y) \\[10pt]
&\hspace{40pt}\rotatebox{45}{$\bigcup$} &           &  \\[-17pt]
&                          &           & \hspace{-20pt}\rotatebox{-45}{$\bigcup$} \\
&                          & \psi_{k}  &
\end{array}
\]

\subsubsection{The Free Algebra $K<Y>$ and its Hopf Algebra Structure}
Let $K<Y>$ be the unital free associative 
$K$-algebra over a field of characteristic
zero (including ${\Bbb Q}$) and
generated by the two-letter alphabet $Y=\{Z^{0}[J], \langle
V(\frac{\delta}{\delta J_{x}})\rangle  _{x}\}$ of non-commuting
variables, with concatenation as multiplication and unit (neutral)
element ${\bf 1}$ the empty word.
Let ${\frak L}_K(Y)$ be the
infinite dimensional free Lie algebra on $Y$ where its elements
are submodules of $K<Y>$ with Lie bracket as multiplication.
$K<Y>$ is the enveloping algebra of ${\frak L}_K(Y)$. We can give
$K<Y>$ a Hopf algebra structure \cite{reut} by defining a
primitive coproduct on the alphabet letters:
\begin{eqnarray}
\Delta({\bf 1})&=&{\bf 1}\otimes {\bf 1},\\
\Delta(Z^{0}[J])&=&{\bf 1}\otimes Z^{0}[J]+ Z^{0}[J]\otimes {\bf 1},\\
\Delta(\langle V\rangle  )&=&{\bf 1}\otimes \langle V\rangle  +
\langle V\rangle  \otimes {\bf 1},\label{ll.1}
\end{eqnarray}
and extending it to words by the connection axiom. The antipode is given by
\begin{eqnarray}
S(a)&=& -a, \;\;\; (a=\langle V\rangle  ,\;\text{or}\; Z^{0}[J])\\
S({\bf 1})&=&{\bf 1},\label{ll.2}
\end{eqnarray}
and is extended to words by the antihomorphism
\begin{equation}
S(a_{1}\dots a_{n})=S(a_{n})\dots S(a_{1}).\label{ll.3}
\end{equation}
The counit map $\varepsilon:K<Y>\rightarrow K$ is defined on the generating letters by
\begin{eqnarray}
\varepsilon(a)&=&0 \;\; (a=\langle V\rangle  ,\;\hbox{or}\; Z^{0}[J])\\
\varepsilon({\bf 1})&=&1,\label{ll.4}
\end{eqnarray}
and is extended to words by the connection axiom.
All the elements (Lie polynomials) $P\in {\frak L}_K(Y)$ are primitives of this Hopf algebra:
\begin{equation}
\Delta(P)={\bf 1}\otimes P+ P\otimes {\bf 1},\label{ll.5}
\end{equation}
and
\begin{eqnarray}
S(P)&=&-P,\\
\varepsilon(P)&=&(P,{\bf 1}), \label{ll.6}
\end{eqnarray}
where $(P,{\bf 1})$ is the coefficient in $P$ of the unit element.\\


We introduce now on $K<Y>$ a derivation  which maps $\langle
V(\frac{\delta}{\delta J_{x}})\rangle  _{x}$ onto 0, and
$Z^{0}[J]$ onto $X$ by means of the operator
\begin{equation}
D=-X\frac{\partial}{\partial Z^{0}[J]}.\label{d.0}
\end{equation}
Clearly $D$ is an endomorphism on $K<Y>$.\\

Since $Z^{0}[J]$ and $\langle V(\frac{\delta}{\delta
J_{x}})\rangle  _{x}$ do not commute, the action of the derivation
$D$ on an arbitrary function of $Z^{0}[J]$ is given by
\cite{reut}:
\begin{equation}
D(f(Z^{0}[J]))=\sum_{k\geq 1} \frac{1}{k!} \text
{ad}(Z^{0}[J])^{k-1}\left(D
Z^{0}[J]\right)f^{(k)}\left(Z^{0}[J]\right), \label{d.1}
\end{equation}
and, in particular,
\begin{eqnarray}
D(\e^{-Z^{0}[J]})&=&\sum_{k\geq 1} \frac{1}{k!} \text {ad}(-Z^{0}[J])^{k-1}(X)\ \e^{-Z^{0}[J]}\nonumber\\
&=& \left(\frac{\e^{\text {ad}(-Z^{0}[J])} -1}{\text {ad}(-Z^{0}[J])}\right)(X)\ \e^{-Z^{0}[J]}\nonumber\\
&=&\e^{-Z^{0}[J]}(X)e^{Z^{0}[J]}\cdot \e^{-Z^{0}[J]}\nonumber\\
&=&-\frac{1}{\lambda}\langle V\left(\frac{\delta}{\delta
J_{x}}\right)\rangle  _{x}\e^{-Z^{0}[J]}, \label{d.2}
\end{eqnarray}
where in going from the third to the last line we have made use of
(\ref{9}). Moreover since $D(\langle V(\frac{\delta}{\delta
J_{x}})\rangle  _{x})=0$, we immediately get
\begin{equation}
\lambda^{n}D^{n}(\e^{-Z^{0}[J]})=\left(-\langle
V\left(\frac{\delta}{\delta J_{x}}\right)\rangle
_{x}\right)^{n}\e^{-Z^{0}[J]}.\label{d.5}
\end{equation}
Using now once more the fact that $D$ is a derivation, we have that the exponential
map $\mu=\e^{\lambda D}$ is a homomorphism of algebras and
\begin{eqnarray}
\mu(\e^{-Z^{0}[J]})&=&\sum_{n\geq 0}\frac{\lambda^{n}}{n!}D^{n}(\e^{-Z^{0}[J]})\nonumber\\
                  &=&\sum_{n\geq 0}\frac{1}{n!}\left(-\langle V\left(\frac{\delta}{\delta J_{x}}
                  \right)\rangle  _{x}\right)^{n}\e^{-Z^{0}[J]}\nonumber\\
                  &=&\e^{-\langle V(\frac{\delta}{\delta J_{x}})\rangle  _{x}}\e^{-Z^{0}[J]}.\label{d.6}
\end{eqnarray}
Furthermore, since $\mu$ is a continuous homomorphism,
\begin{eqnarray}
\e^{-\langle V(\frac{\delta}{\delta J_{x}})\rangle  _{x}}\e^{-Z^{0}[J]}&=&\mu\left(\e^{-Z^{0}[J]}\right)\nonumber\\
&=&\e^{\mu (-Z^{0}[J])}\nonumber\\
&=& \exp\left(\sum_{n\geq
0}\frac{\lambda^{n}}{n!}D^{n}(Z^{0}[J])\right),\label{d.7}
\end{eqnarray}
where
\begin{equation}
D^{n}(Z^{0}[J])=(-1)^{n} \underbrace{\left(X
\frac{\partial}{\partial Z^{0}[J]}\right)\dots\left(X
\frac{\partial}{\partial
Z^{0}[J]}\right)}_{n}(Z^{0}[J]).\label{ll.7}
\end{equation}
Writing
\begin{equation}
\e^{\Psi}=\e^{-\langle V(\frac{\delta}{\delta J_{x}})\rangle
_{x}}\e^{-Z^{0}[J]},\label{ll.8}
\end{equation}
we obtain from (\ref{d.7}) and (\ref{ll.8}) the Hausdorff series relation
\begin{equation}
\Psi= Z^{0}[J]+\sum_{n\geq 1}\frac{\lambda^{n}}{k!}D^{k}(Z^{0}[J]).\label{ll.9}
\end{equation}

Let us now define
\begin{equation}
\Psi_{k}:=\frac{1}{k!} D^{k}(Z^{0}[J]),\label{ll.10}
\end{equation}
which exhibits the operators $\Psi_{k}$ as cyclic vectors with respect to $D$ generated by $Z^{0}[J]$,
and rewrite $X$, introduced in (\ref{9}), as
\begin{equation}
X=-\frac{1}{\lambda}\sum_{j=0}^{d}\frac{1}{j!}[Z^{0}[J],\langle
V\left(\frac{\delta}{\delta J_{x}}\right)\rangle
_{x}]_{j},\label{d.13}
\end{equation}
where $[Z^{0}[J],\langle V(\frac{\delta}{\delta J_{x}})\rangle
_{x}]_{j}$ is defined recursively by $[Z^{0}[J],\langle
V(\frac{\delta}{\delta J_{x}})\rangle
_{x}]_{j}=\break [Z^{0}[J],[Z^{0}[J], \langle V(\frac{\delta}{\delta
J_{x}})\rangle  _{x}]_{j-1}]$, $[Z^{0}[J],\langle
V(\frac{\delta}{\delta J_{x}})\rangle  _{x})]_{0}= \langle
V(\frac{\delta}{\delta J_{x}})\rangle  _{x}$, and the upper index
$d$ in the sum above is the degree of the functional derivative in
$\langle V(\frac{\delta}{\delta J_{x}})\rangle  _{x}$ ($n$=4 for the $\varphi^{4}$ theory).\\

It clearly follows from this and (\ref{ll.7}) that the cyclic vectors $\Psi_{k}$ are elements
of ${\frak L}_K(Y)$ and, hence, primitive elements of the Hopf algebra $K<Y>$.\\

In addition, since $D$ is also an endomorphism on ${\frak L}_K(Y)$ there is a
Hochschild cohomology associated with this algebra for which $D$ is a 1-cochain
$D:{\frak L}_K(Y)\rightarrow {\frak L}_K(Y)$ with coboundary
\begin{equation}
bD(P)=({\text id}\otimes D)\Delta(P)-D(P)\otimes {\bf 1}; \;\; P\in {\frak L}_K(Y).\label{a.252}
\end{equation}

Evidently $bD(P)=0$, because of (\ref{ll.5}),  so the Lie polynomials $P$, and
in particular the $\Psi_{k}$, are 1-cocycles for this cohomology.\\

Note that by applying both sides of (\ref{ll.9}) to the identity function and
recalling (\ref{1}) we recover (\ref{8}), with the functional $\psi_{k}[J]$ given by
\begin{equation}
\psi_{k}[J]=-\frac{1}{k!}D^{k}(Z^{0}[J])(1).\label{d.10}
\end{equation}
Furthermore, since the $\psi_{k}[J]$ are made up by linear combinations of Lie
monomials that do not cancel when acting from the left on the identity they are obviously
also elements of ${\frak L}_K(Y)$ and primitives of the Hopf algebra $K<Y>$.\\


We can derive a recursion relation for the cyclic functionals $\psi_{k}[J]$
by observing that the right side of (\ref{d.10}) can be written as
\begin{equation}
D^{n}(Z^{0}[J])(1)=(-1)^{n}\left( \underbrace{\left(X
\frac{\partial}{\partial Z^{0}[J]}\right)\dots\left(X
\frac{\partial}{\partial
Z^{0}[J]}\right)}_{n-1}X\right)(1).\label{d.10a}
\end{equation}
In particular, for $n=1$
\begin{equation}
-D(Z^{0}[J])(1)=\left(X\frac{\partial}{\partial
Z^{0}[J]}(Z^{0}[J])\right)(1)=X(1)=\psi_{1}, \label{d.11}
\end{equation}
which is in agreement with (\ref{10}) and (\ref{11}).\\

To further see how to interpret the right side of
(\ref{d.10a}) for $n\geq 2$, use (\ref{d.13}) and note that in
order to take into account the fact that the identity cancels
functional derivatives acting on it, we must first apply the $n-1$
derivations in (\ref{d.10a}) to $X$ following the Leibnitz rule,
evaluate each of the resulting derivations on $X$ inside the
commutators by acting on the identity function according to
(\ref{d.10}), and finally act with the resulting bracket
polynomial on the identity. Thus
\begin{equation}
\begin{split}
\psi_{2}&=-\frac{1}{2}\left(\left(X \frac{\partial}{\partial Z^{0}[J]}\right)X\right)(1) \\
&=\frac{1}{2\lambda}\left(X \frac{\partial}{\partial
Z^{0}[J]}\right)\left(\langle V\rangle
-[\langle V\rangle  ,Z^{0}[J]]+\frac{1}{2} [[\langle V\rangle  ,Z^{0}[J]],Z^{0}[J]]\right. \\
&\ \left. + \sum_{j=3}^{d}\frac{1}{j!}[Z^{0}[J],\langle V\rangle  ]_{j}\right)(1) \\
&=\frac{1}{2\lambda}(-[\langle V\rangle
,\psi_{1}]+\frac{1}{2}[[\langle V\rangle  ,\psi_{1}],Z^{0}[J]]
+\frac{1}{2}[[\langle V\rangle  ,Z^{0}[J]],\psi_{1}]+\dots)(1),
\label{ddd.1}
\end{split}
\end{equation}

\begin{equation}
\begin{split}
\psi_{3}&=\frac{1}{6}\left(\left(X \frac{\partial}{\partial
Z^{0}[J]}\right)
\left(X \frac{\partial}{\partial Z^{0}[J]}\right)X\right)(1)                           \\
&= -\frac{1}{6\lambda}\left(X \frac{\partial}{\partial
Z^{0}[J]}\right)\left(-[\langle V\rangle  ,X] +\frac{1}{2}[[\langle V\rangle  ,X],Z^{0}[J]] \right.         \\
&\ \left. +\frac{1}{2}[[\langle V\rangle  ,Z^{0}[J]],X]+\dots\right)(1)               \\
&=-\frac{1}{6\lambda}\left(-\left[\langle V\rangle  ,\left(\left(X
\frac{\partial}{\partial Z^{0}[J]}
\right)X\right)(1)\right]\right)                    \\
&\ +\frac{1}{2}\left[\left[\langle V\rangle  ,\left(\left(X
\frac{\partial}{\partial
Z^{0}[J]}\right)X\right)(1)\right],Z^{0}[J]\right]   \\
&\ +\frac{1}{2}\left[[\langle V\rangle  ,Z^{0}[J]],
\left(\left(X \frac{\partial}{\partial Z^{0}[J]}\right)X\right)(1)\right] 
+[[\langle V\rangle  ,\psi_{1}],\psi_{1}]+\dots \bigg)(1) \\
& =-\frac{1}{6\lambda}\left(2[\langle V\rangle  ,\psi_{2}]-[[\langle V\rangle  ,\psi_{2}],Z^{0}[J]] +[[\langle V\rangle  ,\psi_{1}],\psi_{1}]\right. \\
&\ \left. -[[\langle
V\rangle  ,Z^{0}[J]],\psi_{2}]+\dots\right)(1).\label{ddd.2}
\end{split}
\end{equation}

Iterating on the above, we get the general recursion relation
\begin{equation}
\psi_{n+1}[J]=\frac{1}{n+1}D\psi_{n}[J]=-\frac{1}{n+1}\psi_{1}\frac{\partial}
{\partial Z^{0}[J]}\psi_{n}.\label{d.12}\hspace{.5in}\square
\end{equation}
It should be recalled however that in the implementation of (\ref{d.12}) one has to take
 $\psi_{1}=-\frac{1}{\lambda}\sum_{j=1}^{d}\frac{1}{j!}\left[Z^{0}[J],\langle V(\frac{\delta}
 {\delta J_{x}})\rangle  _{x}\right]_{j}$, perform the derivations to the required order and then
  evaluate on the identity function. \\


We can express the above results in graphical form by making use of Hall trees.
To this end recall \cite{reut} that each Hall tree $h$ of order at least 2 can be written as $h=(h^{\prime},h^{\prime\prime})$, where $h^{\prime}$ and $h^{\prime\prime}$ are the immediate left and right subtrees, respectively, and such that the total ordering 
\begin{equation}
h<h^{\prime\prime}, \ \ h^{\prime}<h^{\prime\prime} \ \ \hbox{and either } 
 h^{\prime}  \in Y, \hbox{ or } h^{\prime}=(x,y) \hbox{ and } y\geqslant h^{\prime\prime}
\end{equation} 
is satisfied. This ordering is lexicographical and is determined by the letters in the alphabet $Y$ that label the leaves. Also, each node in a tree corresponds to a Lie bracket and the foliage $f(h)$ of the tree is the canonical mapping defined by $f(a)=a$ if $a$ is in $Y$ and $f(h)=f(h^{\prime})f(h^{\prime\prime})$ if $h=(h^{\prime},h^{\prime\prime})$ is of degree $\geq 2$.
Now, since a Hall word is the foliage of a unique Hall tree, and since for each Hall word $h$ there is a Lie polynomial $P_{h}$, it can be shown that these Hall polynomials form an infinite dimensional basis of the Lie algebra ${\frak L}_K(Y)$ viewed as a $K$-module.  Parenthetically, one also has that the decreasing products of Hall polynomials $P_{h_1}\dots P_{h_n}$, $h_{1}\geq\dots h_{n}\;$ form a basis of the free associative algebra $K<Y>$.\\
Consequently, using the ordering $\langle  V \rangle <Z^{0}$ for
our two letter alphabet $Y$ and the algorithm given in the proof
of Theorem 4.9 in \cite{reut}, we can always write the $\psi_{i}[J]$'s, 
as derived from (\ref{ll.10}) and (\ref{d.12}), as linear combinations
of Hall polynomials (equivalently Hall trees). Thus, for example 
\vskip -10pt
\begin{equation}
\Psi_1 \equiv X = -\frac{1}{\lambda} \bigg( \langle  V \rangle -
\Tone{Z0}{\langle  V \rangle} +\frac{1}{2}\Ttwo{Z0}{Z0}{\langle
V \rangle} - \frac{1}{6}\Tthree{Z0}{Z0}{Z0}{\langle  V \rangle}
+ \frac{1}{24} \Tfour{Z0}{Z0}{Z0}{Z0}{\langle  V \rangle}
\bigg),\label{f.1}
\end{equation}
so
\begin{equation}
\psi_1 [J] = \Psi_1(1) = -\frac{1}{2\lambda}
\Ttwo{Z0}{Z0}{\langle  V \rangle} +\frac{1}{6\lambda}
\Tthree{Z0}{Z0}{Z0}{\langle  V \rangle} -\frac{1}{24\lambda}
\Tfour{Z0}{Z0}{Z0}{Z0}{\langle  V \rangle};\label{f.2}
\end{equation}

and
\begin{eqnarray}
\Psi_2 &=& -\frac{1}{2\lambda} \Tone{\Psi_1}{\langle  V \rangle}
+\frac{1}{4\lambda} \bigg(
  \Ttwo{Z0}{\Psi_1}{\langle  V \rangle} +
  \Ttwo{\Psi_1}{Z0}{\langle  V \rangle}
\bigg) \nonumber\\ &&
- \frac{1}{12\lambda}
\bigg(
   \Tthree{Z0}{Z0}{\Psi_1}{\langle  V \rangle} +
   \Tthree{Z0}{\Psi_1}{Z0}{\langle  V \rangle} +
   \Tthree{\Psi_1}{Z0}{Z0}{\langle  V \rangle}
\bigg) \nonumber\\ && + \frac{1}{48\lambda} \bigg(
\Tfour{Z0}{Z0}{Z0}{\Psi_1}{\langle  V \rangle} +
\Tfour{Z0}{Z0}{\Psi_1}{Z0}{\langle  V \rangle} +
\Tfour{Z0}{\Psi_1}{Z0}{Z0}{\langle  V \rangle} +
\Tfour{\Psi_1}{Z0}{Z0}{Z0}{\langle  V \rangle}
\bigg),\label{f.3}
\end{eqnarray}

\begin{eqnarray}
\psi_2 [J] &=& -\frac{1}{2\lambda}
\Tone{\hspace{-5pt}\psi_1[J]}{\langle  V \rangle} \hspace{5pt}
+\frac{1}{4\lambda} \bigg(
  \Ttwo{Z0}{\psi_1 [J]}{\langle  V \rangle} +
  \Ttwo{\psi_1[J]}{Z0}{\langle  V \rangle} \hspace{7pt}
\bigg)\nonumber \\ &&
- \frac{1}{12\lambda}
\bigg(
   \Tthree{Z0}{Z0}{\psi_1[J]}{\langle  V \rangle} +
   \Tthree{Z0}{\psi_1[J]}{Z0}{\langle  V \rangle} \hspace{7pt} +
   \Tthree{\psi_1[J]}{Z0}{Z0}{\langle  V \rangle} \hspace{10pt}
\bigg) \nonumber\\ && + \frac{1}{48\lambda} \bigg(
\Tfour{Z0}{Z0}{Z0}{\psi_1 [J]}{\langle  V \rangle} +
\Tfour{Z0}{Z0}{\psi_1[J]}{Z0}{\langle  V \rangle} +
\Tfour{Z0}{\psi_1[J]}{Z0}{Z0}{\langle  V \rangle}\hspace{7pt} +
\Tfour{\psi_1[J]}{Z0}{Z0}{Z0}{\langle  V \rangle}\hspace{10pt}
\bigg).\label{f.4}
\end{eqnarray}
\\
\vskip10pt
Note that (\ref{f.2}) and (\ref{f.4}) for the functionals $\psi_1[J]$ and $\psi_2[J]$
imply acting on the identity function from the left with the diagrams and replacing
in addition  the $\Psi_1$ in the foliage of (\ref{f.3}) by $\psi_1[J]$. The corresponding Hall tree for $\psi_2[J]$ is obtained by grafting (\ref{f.2}) onto each of the branches labeled with $\psi_1[J]$ and implementing the above mentioned algorithm. The procedure is then iterated to whatever order of the $\psi$'s one is interested.\\

By a straightforward calculation one can verify that the final expressions for $\psi_1[J]$ and
$\psi_2[J]$ in terms of propagators for the $\varphi^{4}$ theory are
\begin{equation}
\begin{split}
\psi_{1}[J]& =-\frac{1}{4!}[\langle \Delta_{xa}\Delta_{xb}\Delta_{xc}\Delta_{xd}J_{a}J_{b}J_{c}J_{d}\rangle   \\
&\ -6\langle \Delta_{xx}\Delta_{xa}\Delta_{xb}J_{a}J_{b}\rangle +
3\langle \Delta_{xx}^{2}\rangle  ],\label{12}
\end{split}
\end{equation}

\begin{equation}
\begin{split}
\psi_{2}[J]&=-\frac{1}{2} \langle J_{a}\Delta_{ax} (\frac{1}{6}
\Delta_{xy}^{3}+
 \frac{1}{4}\Delta_{xx}\Delta_{xy}\Delta_{yy})\Delta_{yb}J_{b}\rangle  _{xyab}         \\
&\ -\frac{1}{8} \langle
J_{a}\Delta_{ax}\Delta_{xy}^{2}\Delta_{yy}\Delta_{xb}J_{b}\rangle
_{xyab}
               \\
&\ +\frac{2}{4!}\langle
J_{a}\Delta_{ax}\Delta_{xx}\Delta_{xy}\Delta_{yb}\Delta_{yc}\Delta_{yd}J_{b}
J_{c}J_{d}\rangle  _{xyabcd}\\
&\ +\frac{3}{2(4!)}\langle
J_{a}J_{b}\Delta_{ax}\Delta_{bx}\Delta_{xy}^{2}
\Delta_{yc}\Delta_{yd} J_{c}J_{d}\rangle  _{xyabcd} \\
&\ -\frac{1}{2(3!)^{2}}\langle
J_{a}J_{b}J_{c}\Delta_{xa}\Delta_{xb}\Delta_{xc}\Delta_{xy}\Delta_{yd}
\Delta_{ye}\Delta_{yf}J_{d}J_{e}J_{f}\rangle  _{xyabcdef} \\
&\ + \frac{1}{48}\langle \Delta_{xy}^{4}\rangle  _{xy}  +
\frac{3}{2(4!)}\langle
\Delta_{xx}\Delta_{xy}^{2}\Delta_{yy}\rangle  _{xy},\label{13}
\end{split}
\end{equation}
The expressions for higher order $\psi$'s are increasingly more lengthy, but amenable
to a systematic derivation by the above procedure. This we have done by developing a
REDUCE program which confirms the results given above.\\

\subsubsection{The Hopf algebra $K< \left\lbrace S_{k}\right\rbrace>$}

As we have seen in Eq.(\ref{8}) the power sum symmetric functionals $\psi_{k}[J]$ appear
naturally in PQFT as a result of a perturbative expansion (in the coupling parameter of
the theory)
of the transition amplitude. In terms of the $S_{i}[J]$ these $\psi_{k}[J]$ are
 given  (cf  Eq.(\ref{11}))
by the Schur polynomials. Using as generators the $S_{i}$'s, one can construct a
 universal enveloping
algebra $K< \left\lbrace S_{k}\right\rbrace>$ by introducing a Poincar{\'e}-Birkhoff-Witt basis
$\{1, S_{i_1}, S_{i_1}S_{i_2},
\dots\}$, and defining multiplication $m$ as the disjoint union of the elements of
this basis. Further,
a coalgebra structure can be generated by defining the coproduct \begin{equation}
\Delta(S_{k})=\sum_{i=0}^{k} S_{i}\otimes S_{k-i},\;\;\; S_{0}\equiv {\bf 1},\label{16}
\end{equation}
and a counit $\varepsilon$ as the augmentation of the algebra by
\begin{equation}
\varepsilon(S_{0})=1,\;\; \varepsilon(S_{k})=0, \; k\neq 0. \label{16i}
\end{equation}

We can give this coalgebra the structure of a Hopf algebra by additionally
defining an  antipode $S$ as
 the involutive homomorphism (because of commutativity)
\begin{equation}
S(S_{k})=-S_{k}- m(S\otimes{\text id}){\tilde \Delta}(S_{k}),\label{17}
\end{equation}
where $\tilde{\Delta}$ is the coproduct operation with the primitive contributions removed.\\

Since $K< \left\lbrace S_{k}\right\rbrace>$ is commutative, by the Milnor-Moore
theorem there is a cocommutative Hopf  algebra in duality with it
which is necessarily isomorphic to the universal enveloping
algebra ${\mathcal U}({\frak L})$, where $\frak L$ is a Lie
algebra.
The generators $Z_{i}$ of $\frak L$ are infinitesimal characters
of $K< \left\lbrace S_{k}\right\rbrace>$, {\it i.e} they are linear mappings
$Z_{i}:K< \left\lbrace S_{k}\right\rbrace>\rightarrow K< \left\lbrace S_{k}\right\rbrace>$ fulfilling the conditions :
\begin{eqnarray}
\langle  Z_{i}, S_{k}\rangle &=& \delta_{ik}, \label{19}\\
\langle  Z_{i}, S_{k}S_{l}\rangle&=& \langle  Z_{i}, S_{k}\rangle
\varepsilon(S_{l})+ \varepsilon(S_{k})
 \langle  Z_{i}, S_{l}\rangle.\label{19a}
\end{eqnarray}
We shall denote \cite{kastler} by $\partial \text{Char}K< \left\lbrace S_{k}\right\rbrace>$ the set of
infinitesimal
characters of $K< \left\lbrace S_{k}\right\rbrace>$.
Note that $\frak L$ is Abelian, since the coproduct in $K< \left\lbrace S_{k}\right\rbrace>$ is cocommutative.\\

The exponential mapping $\sum_{i}\alpha_{i} Z_{i}\rightarrow
\text{exp}(\sum_{i}\alpha_{i} Z_{i})
\in{\mathcal G}$, equipped with a convolution product $\ast$ and unit ${\bf 1_{*}}$:
\begin{eqnarray}
\langle  \chi\ast\eta, S_{k}\rangle&=&\langle
\chi\otimes\eta,\Delta S_{k}\rangle,\;\;
\chi,\eta\in {\mathcal G},\label{19b}\\
\langle  {\bf 1_{*}},S_{k}\rangle&=&\varepsilon(S_{k}),\label{19c}
\end{eqnarray}
together with the inverse
\begin{equation}
\chi^{-1}=\chi\circ S, \label{19d}
\end{equation}
generates a subgroup ${\mathcal G}$ of the group of characters of $K< \left\lbrace S_{k}\right\rbrace>$.\\

${\mathcal G}$ is dual to the Hopf algebra $K< \left\lbrace S_{k}\right\rbrace>$ and it is multiplicative,
{\it i.e.}
it satisfies
\begin{equation}
\langle  \chi, S_{k}S_{l}\rangle=\langle  \chi, S_{k}\rangle
\langle  \chi, S_{l}\rangle\;\;\;\;\chi\in{\mathcal G}.\label{20}
\end{equation}

Moreover, from (\ref{19b}) and the fact that the Lie algebra of $\partial
\text{Char}K< \left\lbrace S_{k}\right\rbrace>$
is Abelian, we have that the convolutive product in our case reduces to
\begin{equation}
\e^{\sum_{i}\alpha_{i} Z_{i}} \ast\e^{\sum_{j}\beta_{j}Z_{j}}=\e^{\sum_{i}(\alpha_{i}
+\beta_{i})
 Z_{i}}.\label{20a}
\end{equation}

\subsubsection{The Hopf algebra $K<\{\psi_{k}\}>$}

The Hopf algebra $K< \left\lbrace S_{k}\right\rbrace>$ induces another Hopf algebra
$K<\{\psi_{k}\}>$ by applying a change to normal coordinates
\cite{chryss} to our original Poincar\'e-Birkhoff-Witt basis,
 constructed from the $S_{i}$ coordinates, by means of the map with the canonical element
 ${\bf C}=\e^{\sum_{i} Z_{i}\otimes\psi_{i}}$ which acts as an identity map, {\it i.e.}:
\begin{equation}
\langle  \e^{\sum_{i} Z_{i}\otimes\psi_{i}} , S_{k}\otimes {\text
id }\rangle=S_{k}.\label{20b}
\end{equation}

It is not difficult to verify that the non-linear relation between the $\psi_{k}$'s
and $S_{i}$'s
resulting from (\ref{20b}) is given by (\ref{11}), and that the $\psi_{k}$'s acquire a
primitive coproduct
\begin{equation}
\Delta \psi_{k}= {\bf 1}\otimes\psi_{k}+ \psi_{k}\otimes {\bf 1},\label{17a}
\end{equation}
and an antipode given by
\begin{equation}
S(\psi_{k})=-\psi_{k}.\label{18}
\end{equation}
Note that since (\ref{11}) is invertible $K< \left\lbrace S_{k}\right\rbrace>\simeq K< \left\lbrace \psi_{k}\right\rbrace >$.

As a parenthetical remark we point out that the normal coordinates that we construct here
differ from those recently discussed in \cite{chryss} in relation to the Hopf algebra
of rooted trees, by the fact that in the context of the latter the
comultiplication given by (\ref{17a}) corresponded to non-branched trees,
and that for the more general case of branched trees it was determined by application
of the Baker-Hausdorff-Campbell formula.\\

\subsection{Hopf Primitives and Connected Green Functions}

 In order to establish explicitely the relation of the $\psi_{k}[J]$ functionals to the Green functions,
observe that because the n-legged connected Green functions in PQFT are obtained from the functional variation
\begin{equation}
G_{E}^{(n)}(x_{1},\dots, x_{n})= -\frac{\delta^{n}Z_{E}[J]}{\delta J_{1}\dots \delta J_{n}}
\vline\;_{J=0},\label{14}
\end{equation}
we have that
\begin{equation}
\psi_{k}[J]=\frac{1}{\lambda^{k}}\sum_{n=0} \frac{1}{(n)!}\langle
G_{k}^{(n)}(x_{1}, \dots, x_{n})J_{1}\dots J_{n}\rangle
,\label{14a}
\end{equation}
where $G_{k}^{(n)}$ are the Euclidean Green functions resulting
from adding all the connected Feynman graphs with $k$ vertices and
${n}$ external legs ($n$ is even for the $\varphi^{4}$ theory). Note also that (\ref{14a}) contains
contributions from the Green functions $G_{k}^{(0)}$ which
correspond to vacuum terms. These contributions may be absorbed in
the $\ln W_{0}[0]$
  term in (\ref{8}) via the normalization constant $N$.\\

The analytical expressions for the graphs composing the Green
functions result, in general, in ultraviolet divergences, and the
standard procedure for removing them is to first apply dimensional
regularization and then successively the Forest Formula. It is at
this stage where the Kreimer-Connes Hopf algebra formalism
provides an important insight into the underlying mathematics
behind the Forest Formula of renormalization. Indeed, by noting
that each Feynman diagram corresponds to a decorated rooted tree
(or a sum of decorated rooted trees in the case of overlapping
diagrams) and that these rooted trees, as well as the diagrams
themselves, are the generators of respective universal enveloping
Hopf algebras, these authors have shown that the operation with a
twisted antipode on the algebra provides a systematic application
of the Forest Formula and, consequently, a systematic procedure
for generating the counterterms needed by the theory in order to
cancel the unwanted infinities. Connes and Kreimer further show
\cite{conkre2} the relation between the algebra of characters,
dual to their Hopf algebra, and the Birkhoff algebraic
decomposition. A detailed exposition of these ideas may be found
in the papers cited in the Introduction.\\
Here we only stress once
more the fact that in our approach we deal with the Hopf primitives
$\psi_{k} \in {\frak L}_{K}(Y)$
rather than with Feynman
graphs directly. As we have seen, these Hopf primitives are
Lie polynomials where each monomial is a
Hall tree that itself corresponds to a sum of Feynman diagrams
with $k$-vertices each and an equal number of external legs and loops.
This last observation can be read off directly from the word composed
by the foliage of a Hall tree. First because the number of external legs $E$
in the Feynman diagrams constituting a Hall tree is given by
\begin{equation}
E=2N_{Z^{0}}-N_{\frac{\delta}{\delta J}} V,\label{f.50}
\end{equation}
where
\begin{eqnarray*}
N_{Z^{0}}&=&\text {number of times the letter $Z^{0}$ appears in the word}\\
N_{\frac{\delta}{\delta J}}&=& \text {degree of the functional derivation in the potential}\\
V&=&\text {number of vertices}\\
 &=&\text {number of times the letter $\langle V\rangle  $ appears in the word},
\end{eqnarray*}
so all Feynman diagrams composing a given Hall tree have the same number of external legs.
Second, because the topology of the Feynman diagrams implies that the number of
loops is fixed by the number of vertices and the number of external legs
by the relation
\begin{equation}
(N-2)V=E+2L-2,\label{f.51}
\end{equation}
where $N$ is the total number of legs per vertex ($N$=4 for the $\varphi^{4}$ theory),
and $L$ is the number of loops, so also all Feynman diagrams composing a Hall tree have the same loop number. \\

\section{Discussion}
As already mentioned in the Introduction, given that a theory is renormalizable, the Hopf Algebra of Renormalization developed by Connes and Kreimer provides insight into the mathematical structures behind the Forest Formula and is extremely useful in numerical calculations since it gives a systematic procedure (amenable to computer programming \cite{broad}) for evaluating the renormalized Green functions. In a parallel vein, the Hopf algebra $K<Y>$ investigated here, helps to exhibit some complementary  mathematical structures associated with the Hausdorff series expansion of perturbation theory.

We have shown the relation of the $\psi_{k}[J]$'s to the Hall polynomials, and that of their graphic representation in terms of Hall trees to the Feynman diagrams. Also, making use of the  cyclic vector character of the $\psi_{k}[J]$'s with respect to the derivation endomorphism $D$, which appears as part of the structure of the Lie subalgebra of $K<Y>$, we have obtained a systematic procedure (also subject to computer programming) for constructing all the Green functions of the theory starting from the generator $Z^{0}[J]$. Moreover, since each action of $D$ introduces an additional vertex in the commutators conforming the $\psi_{k}[J]$'s, one could hope that a further study of this operator and its possible deformations would result in further mathematical insights on the process of renormalization. \\

Note, however, that because of the Hopf primitive character of our $\psi_{k}[J]$'s and their relation to the Green functions given by (\ref{14a}) one should not expect that the application of the algebraic Birkhoff decomposition to our Hopf algebra should be immediately related to the Forest Formula. The Hopf Algebra of Renormalization begins where our Hopf algebra ends. Nonetheless, a unital $K$-algebra homomorphism
 $\phi:\{ K<Y>\} \rightarrow{\bf A}$ from the free algebra $ K<Y>$ to the (unital) $K$-algebra ${\bf A}$ of meromorphic functions
on the Riemann sphere with poles at the origin, followed by a the Birkhoff decomposition allows us to exhibit more explicitely and in a heuristic fashion the point made in the second paragraph of the Introduction. Indeed \cite{kastler}, if we let ${\bf A}={\bf A}_{-}\oplus {\bf A}_{+}$ be a
Birkhoff sum of the $K$-linear multiplicative subspaces ${\bf
A}_{-}$ and  ${\bf A}_{+}$, where ${\bf A}_{-}= \{\text
{polynomials in $z^{-1}$ without constant}$ $\text{term}\}$ and
${\bf A}_{+}= \{\text {Restriction to $(\Bbb C -\{0\})$ of
functions in ${\bf Holom}(\Bbb C)$}\}$, and if we further let $T: {\bf
A}\rightarrow {\bf A}_{-}$ be the Rota-Baxter projection operator, satisfying the multiplicative constraints
\begin{equation}
T(ab) + (Ta)(Tb) = T[(Ta)b + a(Tb)], \; a,b\in {\bf A},\label{23a}
\end{equation}
then the algebraic
 Birkhoff decomposition
\begin{equation}
\phi_{+}=\phi_{-}\star \phi,\label{25f}
\end{equation}
(where the operator $\star$ denotes the convolution product $(\phi \star \phi^{\prime})(w)=m_{{\bf
A}}(\phi\otimes\phi^{\prime})(\Delta w), \;\phi,\phi^{\prime}\in
Hom_{K-alg}(K<Y>,{\bf A})$), together with the Hopf primitive character of the $\psi_{k}[J]$'s immediately imply that
\begin{equation}
\phi(\psi_{k}[J])=\phi_{+}(\psi_{k}[J])-\phi_{-}(\psi_{k}[J])=\phi_{+}(\psi_{k}[J])+T\phi(\psi_{k}[J]).\label{25g}
\end{equation}
 
Thus, by virtue of (\ref{14a}),
\begin{equation}
\begin{split}
\phi_{+}(\psi_{k})=\frac{1}{\lambda^{k}}\sum_{n\geq 1}
\frac{1}{(n)!}\left(\langle  G_{k}^{(n)}(x_{1},
\dots, x_{n})J_{1}\dots J_{n} \rangle \right. \\
-\langle T[G_{k}^{(n)}(x_{1},\dots, x_{n})]J_{1}\dots
J_{n} \rangle\Big),\label{28bb}
\end{split}
\end{equation}
where we have put the projector $T$ inside of the integration in the second
term of (\ref{28bb}) after taking into account that the currents $J_{i}$ are
good test functions, so the pole structure of the integrals is determined
by that of the Green functions.
In addition, the projection by $T$ to $\bf {A}_{-}$ of the connected Green
functions is taken in the Mass Independent Renormalization Scheme.\\

Recall now the basic equation of the renormalization group which relates the dimensionally regularized bare and renormalized connected Green functions, and which in our notation reads:
\begin{equation}
\frac{\partial}{\partial J_{1}}\dots \frac{\partial}{\partial J_{n}}\sum_{k \geq 0}(\lambda_{b})^{k}(\psi_{k}[J])_{b}=\frac{\partial}{\partial J_{1}}\dots \frac{\partial}{\partial J_{n}}\sum_{k\geq 0}\lambda^{k} (\psi_{k}[J])_{R}.\label{28bbb}
\end{equation}
In the above, the subscript $b$ denotes that the regularized Green functions in the momentum representation, occurring in the  $\psi_{k}[J]$'s, are expressed in terms of the bare parameters of the theory, {\it i.e.}  $(\psi_{k}[J])_{b}=\sum_{n=0}(\psi^{(n)}_{k}[J])_{b}=\frac{1}{\lambda_{b}^{k}}\sum_{n=0} \frac{1}{(n)!}\langle
(G^{(n)}_{k})_{b}(p_{1},\dots, p_{n};\lambda_{b}, m_{b},
\epsilon)$, $\lambda_{b}=Z_{\lambda} \lambda$, $m_{b}^{2}=Z_{m}m^{2}$,  $(J_{k})_{b}=Z_{\varphi}^{-1/2} J_{k}$, while the subscript $R$ on the right side of (\ref{28bbb}) denotes the $\psi_{k}$'s evaluated with the renormalized Green functions. Furthermore, in the Mass Independent Scheme the counterterms contain no finite contributions and
are polynomials in inverse powers of $\epsilon$ with coefficients depending only on $\lambda$,
so the renormalization parameters $Z_{\lambda}$, $Z_{\varphi}$ and  $Z_{m}$ are polynomials in inverse powers of $\epsilon$ with coefficients depending only on $\lambda$.\\ 

If we replace (\ref{25g}) in (\ref{28bbb}) and observe that the right side of the later equation is finite, we thus arrive at the following set of equations:
\begin{equation}
\begin{split}
\lim_{\epsilon\rightarrow 0}[{\rm Finite}\{\sum_{k \geq 0}(Z_{\lambda})^{k}Z_{\varphi}^{-\frac{n}{2}}(G^{(n)}_{k}(p_{1},\dots, p_{n};\lambda, Z_{m}m^{2},
\epsilon))_{(+)}\}]=\\
\sum_{k\geq 0}(G^{(n)}_{k}(p_{1},\dots, p_{n};\lambda, m,
)_{R},\label{29bb}
\end{split}
\end{equation}
and
\begin{equation}
{\rm Poles}\{\sum_{k \geq 0}(Z_{\lambda})^{k}Z_{\varphi}^{-\frac{n}{2}}(G^{(n)}_{k}(p_{1},\dots, p_{n};\lambda, Z_{m}m^{2},
\epsilon))_{(-)}\}=0, \label{29cc}
\end{equation}

where
\begin{equation}
\begin{split}
(G^{(n)}_{k}(p_{1},\dots, p_{n};\lambda, Z_{m}m^{2},
\epsilon))_{(+)}:=\\
[G_{k}^{(n)}(p_{1},\dots, p_{n},Z_{m}m^{2},\epsilon)
 -T(G_{k}^{(n)}(p_{1},\dots, p_{n},Z_{m}m^{2},\epsilon)],\label{30bb}
\end{split}
\end{equation}
and
\begin{equation}
(G^{(n)}_{k}(p_{1},\dots, p_{n};\lambda, Z_{m}m^{2},
\epsilon))_{(-)}:=T(G_{k}^{(n)}(p_{1},\dots, p_{n},Z_{m}m^{2},\epsilon).\label{31bb}
\end{equation}

But if, and only if, the theory is renormalizable the equations (\ref{29cc}) can be solved consistently and order by order in the powers of $\lambda$, ${\it i.e.}$ only then the counterterms can be absorbed into the parameters of the theory, and only then the theory will be physical. Formally, the substitution of the resulting renormalization functions into the left side of (\ref{29bb}) would then yield the physically meaningful renormalized Green functions. In such a case, however, the standard recursive elimination of subdivergences via the Forest Formula (or the Connes-Kreimer Hopf algebra) would clearly be a more efficient manner to derive these quantities. On the other hand, the opposite is not necessarily true: The fact that the forest formula provides a finite answer does not suffice to make the theory physical, it is also necessary that the poles of the ill-defined regularized Green functions can be absorbed into the bare parameters of the theory.\\

To conclude, we would like to comment on several other possible lines for future work in addition to the ones already mentioned. Thus, for example, since there is a duality between concatenation and shuffle products \cite{reut, sweedler}, there are actually two bialgebra structures on $ K<Y>$: The one considered here with product (words) made out by concatenation of letters of the alphabet $Y$ and coproduct defined by (16-18), the other one with shuffle product and coproduct defined by
\begin{equation}
\Delta^{\prime}(w)=\sum_{u,v\in Y^{*}}(w,uv)u\otimes v,\label{32bb}  
\end{equation}
where $Y^{*}$ is the free monoid on $Y$. It may be interesting to investigate the relation of this other bialgebra to the primitives $\psi_{k}[J]$ discussed here, and if this relation has  
a possible relevance to renormalization.\\
One could also ask if it is possible to carry out a program starting from the Hopf algebra of Feynman diagrams and perform a Legendre transformation to a Hopf algebra of $\psi_{k}$, which would no longer be primitive, and by exponentiation to a Hopf algebra of $S_{k}$. It would be interesting then to consider not only the Hopf algebra $K\{\psi_{k}\}$ but also 
$K\{\psi_{k}\}/\sim$, where $\sim$ stands for an equivalence relation coming ${\it e.g.}$ from a Ward identity (we are grateful to the referee for pointing out this possible program).
Some of these lines of work are under present consideration and the results will be reported elsewhere.

\section{Acknowledgment} The authors are grateful to Prof. Dirk Kreimer and to the referee for valuable comments and suggestions regarding this work and recommendations leading to the improvement of the presentation. We are also grateful to Dr. J.C. L{\'o}pez-Vieyra for help with the graphs. We acknowledge partial support from CONACYT project G245427-E (M.R.), DGAPA-UNAM grant IN112401 (H.Q.), and DGAPA-UNAM grant IN117000 (J.D.V.).

\end{document}